\documentclass[letterpaper]{article}

\usepackage[T1]{fontenc}

\usepackage{geometry}
\usepackage{setspace}

\usepackage{achemso}

\usepackage{graphicx}
\usepackage{float}
\newfloat{scheme}{htbp}{los}
\floatname{scheme}{Scheme}
\floatname{chart}{Chart}
\newfloat{graph}{htbp}{loh}

\usepackage{chemformula} 
\usepackage[version = 4]{mhchem} 

\usepackage{authblk}
\author[1]{Yi-Fan Yang$^{\#}$}
\author[1]{Di Liu$^{\#}$}
\author[1]{Zhong-Hua Cui$\star$}
\author[1]{Bing Yan$\dag$}
\affil[1]{Institute of Atomic and Molecular Physics, Jilin University, Changchun 130023, China}
\author[2]{Lorenz S. Cederbaum$\ddag$}
\affil[2]{Theoretical Chemistry, Institute of Physical Chemistry, Im Neuenheimer Feld 229, Universit{\"a}t Heidelberg, D-69120 Heidelberg, Germany}

\title{Intriguing Electronic Structures of {C}$_{8}$ and {C}$_{12}$ Carbon Rings}

\date{$\star$Email: zcui@jlu.edu.cn\\$\dag$Email:yanbing@jlu.edu.cn\\$\ddag$Email:Lorenz.Cederbaum@pci.uni-heidelberg.de}

\begin{document}

\maketitle

\def\thefootnote{{\#}}\footnotetext{These authors contributed equally to this work}\def\thefootnote{\arabic{footnote}}

\begin{abstract}
Since the success in synthesizing and characterizing the first carbon ring {C}$_{18}$ in 2019, they have attracted great attention both experimentally and computationally. In this work we investigate the electronic and geometric structures of the {C}$_{4n}$ carbon rings {C}$_{8}$ and {C}$_{12}$ by employing state-of-the-art coupled cluster and multi-reference configuration interaction methods. Known is that {C}$_{12}$ has been synthesized and characterized and an experiment on {C}$_{8}$ has been performed. Computationally, only the ground states of these rings have been reported in the literature. We report on the ground and numerous excited electronic states. In the ground state the {C}$_{4n}$ rings are closed-shell systems possessing polyynic structures and can be classified as double anti-aromatic molecules. In their energetically lowest lying triplet state the rings exhibit aromatic cumulenic structures.  The overall change in the electronic structures is rather dramatic upon the found moderate geometric changes from polyynic to cumulenic structure. Among others, Hund{'}s rule is violated in both {C}$_{8}$ and {C}$_{12}$ in their cumulenic structures. We mention that until now, graphene is the only carbon allotrope reported to violate Hund{'}s rule. The reasons for the violation are analyzed. Much effort has been invested to understand the relaxation pathways of the low-lying states leading the {C}$_{8}$ from polyynic to cumulenic geometry and vice versa.  On its minimum energy path, the first singlet excited state changes from open-shell character in the polyynic structure to a closed-shell state in the cumulenic structure. The cumulenic state lowest in energy is an open-shell singlet which relaxes to the closed-shell polyynic ground state. The lowest triplet polyynic state relaxes to a stable cumulenic structure energetically placed between these two singlets. The found states and interconnecting pathways allow us to construct a scheme to arrive at the cumulenic structure starting from the polyynic ground state of the ring. 
\end{abstract}

\section*{Keywords}

double aromaticity, carbon monocyclic rings, disjoint diradical, vibronic coupling, minimum energy path

\section{Introduction}

Monocyclic carbon rings are new members of the carbon allotrope family. Unlike fullerenes\cite{guldi2000excited}, graphenes,\cite{allen2010honeycomb} carbon nanotubes,\cite{tasis2006chemistry} the synthesis and characterization of carbon rings pose great challenges to experimental researchers.\cite{anderson2021short} For decades, only a few experimental studies were reported.\cite{prinzbach2000gas,belau2007ionization} In 2019, Anderson{'}s group overcame the underlying experimental bottleneck and characterized a fascinating carbon ring ({C}$_{18}$) by employing the state-of-the-art Scanning Tunneling Microscope (STM) method.\cite{kaiser2019sp}  Inspired by this breakthrough, between 2023 and 2025, researchers synthesized and characterized all even number monocyclic carbon rings between {C}$_{10}$ and {C}$_{20}$.\cite{gao2023surface,sun2023surface,sun2024surface,gao2024surface,sun2025surface} Carbon rings have also attracted much attention from the theoretical community. Various investigations predicted that carbon rings may have great application potentials in many fields, e.g., optical nonlinearity,\cite{liu2020sp}, optical molecular switches,\cite{liu2022potential} storage of metal atoms,\cite{yang2021endocircular,yang2022storing} and as stable doubly-excited anions.\cite{hou2023can} Monocyclic carbon rings are fascinating materials with new properties and opportunities.\cite{liu2025cyclo}

Carbon rings are so-called \verb+"+double aromatic\verb+"+ species, which possess two types of delocalized $\pi$ orbitals (one in plane and one vertical to the plane),\cite{doublearomaticity} rather than only one delocalized set like in benzene.\cite{Rosi_2004,baldea2007jahn} This \verb+"+double aromatic\verb+"+ concept was first introduced by Dewar\cite{dewar1979sigma} and named by Schleyer and his co-workers.\cite{doublearomaticity} Naturally, researchers are interested in the aromaticity of these double aromatic species. The popular H\"uckel{'}s rule predicts that the ground states of {C}$_{4n+2}$ rings, e.g., {C}$_{18}$, are aromatic and the ground states of {C}$_{4n}$ rings are anti-aromatic. However, for decades, the aromaticity issue of {C}$_{18}$ rings could not be settled, since different theoretical methods provided contradicting predictions. Most density functional theory (DFT) and second{-}order M{\o}ller{-}Plesset perturbation theory (MP2) calculations supported the prediction of aromatic {C}$_{18}$ ring and that the geometry of its ground state is cumulenic.\cite{kaiser2019sp} In contrast, coupled cluster and high-level Monte Carlo methodologies predict anti-aromaticity and polyynic geometries.\cite{PhysRevLett.85.1702,arulmozhiraja2008ccsd} This chaos of predictions on {C}$_{18}$ lasts until 2019, when the results of STM experiment supported the polyynic structure prediction.\cite{kaiser2019sp} It indicates that accurately describing carbon rings is still an ongoing challenge to popular electronic structure methods, e.g., DFT.\cite{bremond2019challenge} 

Due to the complex nature of the carbon rings, it is necessary to compare the results of highly accurate theoretical methods and predictions of popular electronic structure rules. In the literature, typical {C}$_{2n}$ rings possess closed-shell ground states and open-shell excited states,\cite{anderson2021short} whose properties are predicted by H\"uckel{'}s rule\cite{anderson2021short} and Baird{'}s rule,\cite{baird1972quantum, JORNER2021375} respectively. Since there are exceptions of H\"uckel{'}s rule in carbon rings, it is also interesting to study the prediction of Baird{'}s rule on carbon rings. Baird{'}s rule predicts that the open-shell excited states of {C}$_{8}$ and {C}$_{12}$ with two unpaired electrons are aromatic. The reason is that these two electrons occupy the non-bonding orbitals and the doubly-occupied orbitals can be treated as a {C}$_{4n-2}$ system.\cite{JORNER2021375} However, little attention has been paid to the excited states of carbon rings until now and the aromaticity of the excited states of {C}$_{2n}$ rings is still unknown. Here, further studies with state-of-the-art methods are required.

A typical aromatic molecule possesses a closed-shell ground state and several of its low lying excited states are open shell. If the two singly occupied natural orbitals (SONOs) of excited states of molecules are degenerate or quasi-degenerate, these molecules are diradicals.\cite{stuyver2019diradicals} In the literature, the aromaticity is the driving force for the preference of a diradical structure.\cite{stuyver2019diradicals} Thus, the stable diradicals may be aromatic, due to this stabilizing effect. In addition, there is a special class of diradicals, i.e., disjoint diradicals, which until now has been found only in hydrocarbons and their derivatives.\cite{dias2003disjoint,zhang2019stable,moles2022polycyclic} In disjoint diradicals, two radical units occupy disjoint sets of atoms, and the SONOs possess nodes on the connecting atoms.\cite{trinquier2024search} Due to the disjoint nature, there is no interaction between the two radical units, resulting in zero or small so-called kinetic exchange.\cite{trinquier2024search} In addition, if the two unpaired electrons occupy disjoint non-bonding molecular orbitals, the triplet and singlet open-shell states are nearly degenerate at Hartree-Fock level.\cite{dias2003disjoint} Thus, one of the fundamental electronic structure rules, Hund{'}s multiplicity rule, may be violated. Such materials have great application potential in organic optoelectronic devices (OLED).\cite{sobolewski2023excited, sanz2021negative, aizawa2022delayed, pollice2021organic, borden1994violations} In the literature, graphene is the only carbon allotrope reported to violate Hund{'}s rule.\cite{sheng2013violation} Thus, studying the excited states of carbon rings may help us design promising materials with interesting properties.

In this work we systematically investigate the electronic and geometric structures of the {C}$_{8}$ and {C}$_{12}$ rings by employing {\textit {ab initio}} state-of-the-art coupled cluster\cite{CCSD, stanton1993equation, levchenko2004equation} and multi-reference configuration interactions methods.\cite{werner1988efficient,knowles1988efficient,knowles1992internally} The ground and all the low-lying excited electronic states are computed at their polyynic as well as at their cumulenic structures. Particular attention is paid in the example of {C}$_{8}$ to the minimum energy pathways connecting these structures and to how they can be utilized to arrive to the cumulenic structure starting from the polyynic ground state of the ring.  We find that both {C}$_{8}$ and {C}$_{12}$ possess all the aforementioned intriguing properties, i.e., Baird{'}s aromaticity, stable diradicals, disjoint SONOs, and violation of Hund{'}s rule. This is the first report of violation of Hund{'}s rule in carbon allotropes aside from graphenes, and the first report of this violation in double aromatic molecules. The investigation shows that carbon rings are fascinating molecules with properties which can be exploited.  

\section{Results and discussion}

In this section we first present our results for the ground and low-lying excited electronic states followed by a discussion of these results. Thereby, we pay particular attention to the minimum energy pathways between the polyynic and cumulenic structures, to the violation of Hund{'}s rule and to the symmetry breaking mechanism in the lowest energy state of cumulenic structure.

\subsection* {\label{Geo} 2.1 Ground states of polyynic and cumulenic structures of {C}$_{8}$ and {C}$_{12}$ rings}

We optimized the geometries of the {C}$_{8}$ and {C}$_{12}$ rings employing coupled cluster methods and cc-pVTZ basis sets. The global minima have been found to belong to closed-shell states. The resulting respective coordinates are shown in section S1 of the supporting information (SI). The equilibrium geometries are shown in the Figure 1. H\"uckel{'}s rule predicts anti-aromaticity for {C}$_{4n}$ rings\cite{anderson2021short} and, hence, the geometrical structures of these rings in their global ground states are expected to be polyynic (unequal C-C bond lengths). This prediction is consistent with our coupled cluster calculations and the experimental characterization results of {C}$_{12}$.\cite{sun2024surface} 

We have also computed the cumulenic structures where all C-C bond lengths are equal. Calculating these structures is much less straight forward as we have found the cumulenic closed-shell state to be higher in energy than the cumulenic singlet and triplet states lowest in energy. Here, the two unpaired electrons occupy two quasi-degenerate orbitals making the system a diradical. Unfortunately, the traditional equation of motion coupled cluster method is not suitable to study diradicals, as the restricted Hartree-Fock reference wavefunctions is unstable.\cite{krylov2006spin} Consequently, we turn to the equation-of-motion spin-flip coupled cluster (EOM-SF-CCSD)\cite{levchenko2004equation} method to optimize the geometry of cumulenic structure, as has been successfully done in the studies of several diradical molecules.\cite{krylov2006spin} The optimized cumulenic structures of {C}$_{8}$ and {C}$_{12}$ are depicted in Figure 1 and the respective coordinates are collected in section S1 of the SI. 

The vibrational frequencies at the optimized structures have also been calculated and reported in section S2 of the SI. The vibrational frequencies show that the polyynic structure is a minimum on the potential energy surface (PES). As discussed in section S1 of the SI, three parameters are sufficient to describe the structures of carbon rings. They are an angel $\theta$$_{1}$, and two radii {R}$_{1}$ and {R}$_{2}$. To our surprise, the differences in the geometries of the optimized polyynic and cumulenic structures are rather small. In {C}$_{8}$, angle $\theta$$_{1}$ changes by only 2.7 degrees, when the structure turns from polyynic to cumulenic, see Figure 1. Similarly, the change of radii is around 0.01 $\AA$ and 0.02 $\AA$. The respective changes in {C}$_{12}$ are 2 degrees for $\theta$$_{1}$ and 0.04 $\AA$ and 0.16 $\AA$ for the radii.

The resulting vibrational frequencies make clear that not only the closed-shell polyynic ground states are minima on the respective PES of {C}$_{8}$ and {C}$_{12}$. They show that the lowest triplet states of the {C}$_{8}$ and {C}$_{12}$ rings are also minima on the respective PES, however, at cumulenic geometries. The cumulenic structure of the triplet states indicates aromaticity and representing a new example of application of Baird{'}s rule in carbon rings.

\begin{center}
\includegraphics[scale=0.6]{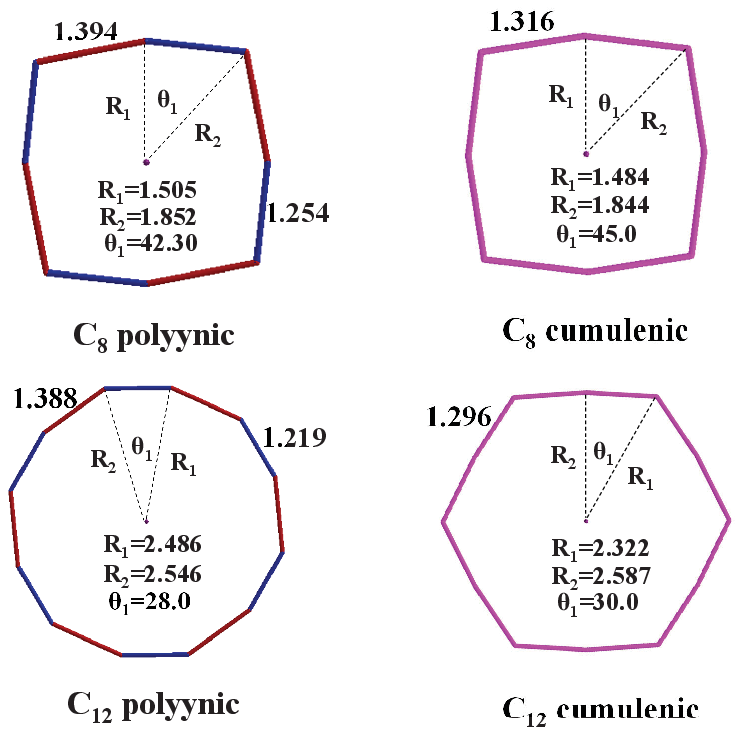}
\end{center}
\begin{flushleft}
{\bf Figure 1. The computed geometries of polyynic structures of the ground states and cumulenic structures of lowest triplet excited states of {C}$_{8}$ and {C}$_{12}$ rings. The ground states of polyynic structures of {C}$_{8}$ and {C}$_{12}$ are optimized at the CCSD/cc-pVTZ level. The triplet excited states are optimized at EOM-SF-CCSD/cc-pVTZ level. The equivalent bonds are indicated by the same color. Three parameters suffice to describe the structures of carbon rings, i.e., an angel $\theta$$_{1}$, and two radii {R}$_{1}$ and {R}$_{2}$, which are shown in figure. The unit of bond length is $\AA$ and the unit of angle is degree.}
\end{flushleft}

As mentioned above, for both {C}$_{8}$ and {C}$_{12}$ at the cumulenic geometry, the triplet state $^{3}${A}$_{2g}$ is lower in energy than the first closed-shell state $^{1}${A}$_{1g}$. As we will discuss further in the following paragraphs, in the cumulenic structure these stable triplet states are not the states lowest in energy as one would expect from Hund{'}s rule. In both {C}$_{8}$ and {C}$_{12}$, the partner open-shell singlet states are the cumulenic ground states. This interesting violation of Hund{'}s rule will be analyzed in detail below. In contrast to the triplet state which is a true minimum on the PES, the respective singlet open-shell state does not possess a minimum at the cumulenic geometry. It can be seen from section S2 of the SI that the singlet state $^{1}${A}$_{2g}$ of {C}$_{8}$ exhibits a single imaginary vibrational frequency. In other words, the singlet open-shell state at the cumulenic geometry is on the one hand side the state lowest in energy and on the other hand a transition state.\cite{fueno2019transition} Where this transition state will lead to when occupied will be discussed below.

\subsection* {\label{excited}2.2 Low-lying excited states and violation of Hund{'}s rule}

To have a better understanding of the excited states of the carbon rings, we calculated the vertical excitation energies of the low-lying electronic states based on the optimized polyynic and cumulenic structures discussed in the former section. All these states are singly excited and can be characterized by two singly occupied natural orbitals (SONOs). For simplicity, we only show the low-lying excited states of {C}$_{8}$ here. The states with relative energies lower than 3 eV are listed in Figure 2. All the computed states of {C}$_{8}$ and {C}$_{12}$ are shown in sections S3 and S4 of the SI, respectively.

\begin{center}
\includegraphics[scale=0.55]{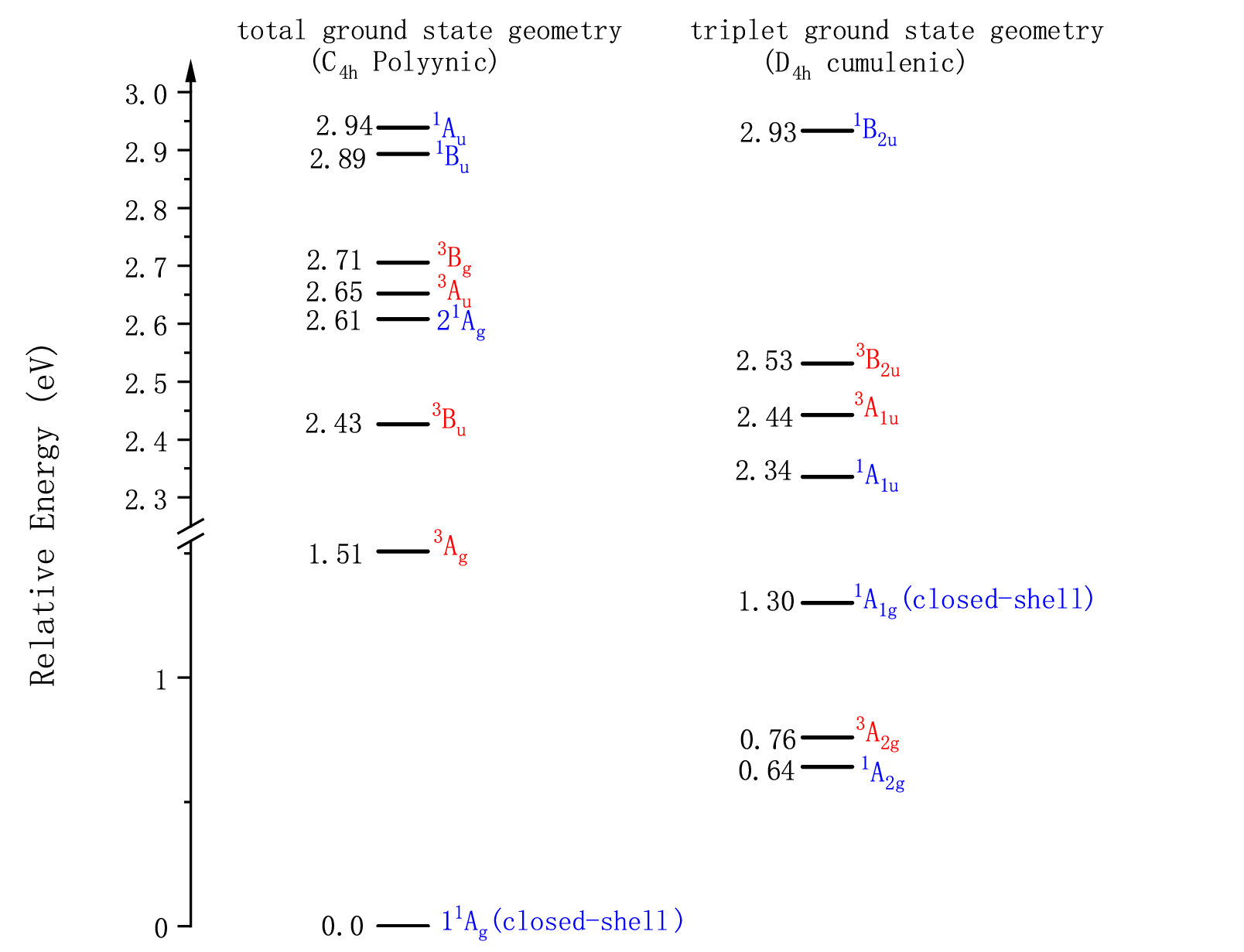}
\end{center}
\begin{flushleft}
{\bf Figure 2. Vertical excitation energies of the low-lying excited states of {C}$_{8}$ in polyynic structure ({C}$_{4h}$ symmetry) and cumulenic structure ({D}$_{4h}$ symmetry) obtained at the EE-EOM-CCSD/cc-pVTZ level. The geometries are the equilibrium geometries shown in Figure 1. The singlet and triplet states are marked in blue and red, respectively. It is noteworthy that in the cumulenic structure the closed-shell state of cumulenic structure is an excited state. The unit of energy is in eV. More excited states of {C}$_{8}$ and those of {C}$_{12}$ are collected in section S3 and S4 of the Supporting Information.}
\end{flushleft}

As shown in Figure 2, all the open-shell states of polyynic structure are higher in energy than the closed-shell state, which is the true ground state of {C}$_{8}$. It is noteworthy that there is a large gap (1.5 eV) between the ground state and the first excited state in polyynic structure. All the singlet open-shell states are higher than the corresponding triplet states in energy in agreement with Hund{'}s rule.

The cumulenic {C}$_{8}$ ring possesses an interesting electronic structure. The EE-EOM-CCSD calculations show that the lowest closed-shell state is an excited state in the cumulenic structure. In addition, the lowest singlet open-shell state $^{1}${A}$_{2g}$ is lower in energy than its triplet state counterpart $^{3}${A}$_{2g}$, which violates Hund{'}s rule. This result is consistent with our results obtained at the EOM-SF-CCSD/cc-pVTZ level. To study the mechanism leading to this violation, we will have a closer look at the frontier orbitals and potential energy surface in the following section. Similarly, we calculated the excitation energies of singly excited states of {C}$_{12}$ at the EE-EOM-CCSD/cc-pVTZ level based on the equilibrium geometries. The results are shown in section S3 of the SI. We found that, similar to the {C}$_{8}$ ring, Hund{'}s rule is also violated in the cumulenic structure of {C}$_{12}$. This indicates that all {C}$_{4n}$ ring may have similar properties and, accordingly, possess application potential.

\subsection* {\label{orbitals}2.3 Visualization of the singly occupied natural orbitals of the electronic states of the {C}$_{8}$ ring}

Zilberg and Haas\cite{zilberg1998two} suggested that triplet Baird{'}s aromatic molecules with 4n $\pi$-electrons can be seen as H\"uckel systems with 4n-2 $\pi$-electrons. In addition, the two unpaired electrons do not contribute to the bonding.\cite{zilberg1998two} Thus, to have a better understanding of the mechanism of aforementioned properties, we visualized the singly occupied natural orbitals (SONOs) of lowest open-shell states of {C}$_{8}$ in both polyynic and cumulenic structures, i.e., $^{3}${A}$_{g}$ and $^{1}${A}$_{2g}$ states, respectively, see Figure 3.

As one can see in Figure 3, the electrons of the SONOs of the $^{3}${A}$_{g}$ state of polyynic structure calculated at EOM-CCSD/cc-pVTZ level are distributed on the C-C bonds. The $^{3}${A}$_{g}$ state is a typical example of open-shell states in carbon rings, as most open-shell states share the distribution pattern of these SONOs, see section S3 of the SI, where we have collected all SONOs of the computed open-shell states.

\begin{center}
\includegraphics[scale=0.4]{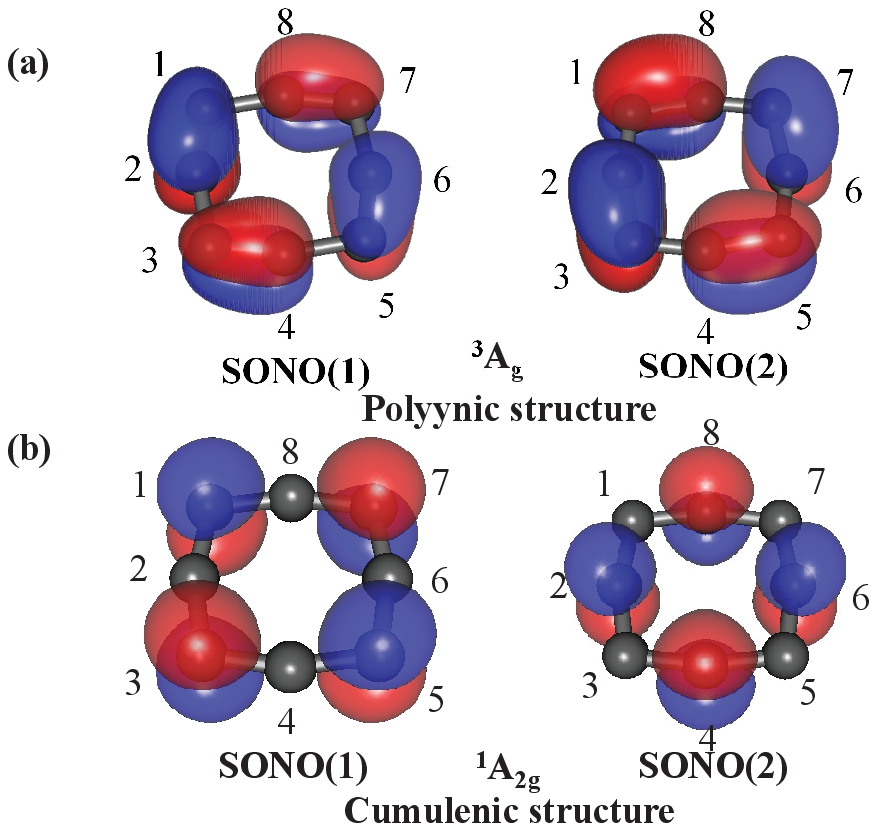}
\end{center}
\begin{flushleft}
{\bf Figure 3. Images of the singly occupied natural orbitals (SONOs) of the lowest open-shell state in the polyynic structure and in the cumulenic structure of the {C}$_{8}$ ring at EOM-CCSD/cc-pVTZ level{:} (a) $^{3}${A}$_{g}$ state of polyynic structure (b) $^{1}${A}$_{2g}$ state of cumulenic structure. The structures are depicted in Figure 1. The surfaces shown enclose 80$\%$ of the orbital density. The numbers shown enumerate the carbon atoms. Obviously, the electron clouds of the SONOs of $^{3}${A}$_{g}$ state is distributed on the C-C bonds{:} SONO(1) on the bonds between the carbon atom pairs 1-2, 3-4, 5-6 and 7-8, while SONO(2) on the bonds between the complementary carbon atom pairs 2-3, 4-5, 6-7 and 8-1. In strong contrast to this behavior, the unpaired electrons of the lowest open-shell state $^{1}${A}$_{2g}$ in the cumulenic structure are distributed over individual carbon atoms, SONO(1) on the odd carbon atoms and SONO(2) on the even ones, and clearly do not contribute to the bonding. Clearly, the two unpaired electrons of the $^{1}${A}$_{2g}$ state do not contribute to bonding, unlike the unpaired electrons of $^{3}${A}$_{g}$. This may be the reason for the aromaticity in the cumulenic structure, as discussed in the main text. These findings are characteristic to the open-shell states, see section S3 of the SI.}
\end{flushleft}

The unique properties of the $^{1}${A}$_{2g}$ state of cumulenic structure can be related to its SONOs. SONO(1) and SONO(2) are disjoint, the respective electron clouds being distributed on alternating odd and even carbon site numbers, respectively. Naturally, one may expect that these two SONOs are degenerate. The irreducible representations of SONO(1) and SONO(2) are {B}$_{2u}$ and {B}$_{1u}$, respectively, so that there is no symmetry induced degeneracy. However, the two orbitals are quasi-degenerate with an energy gap of only 0.7 eV at the EOM-SF-CCSD/cc-pVTZ level. The corresponding triplet state has similar disjoint SONOs. It is noteworthy that the cumulenic triplet state is stable (no imaginary frequencies). Thus, it is the first reported example of a stable disjoint diradical in carbon allotropes.

Due to the non-bonding nature of the SONOs, the low-lying open-shell cumulenic states possess aromaticity. The results are collected in section S3 of the SI. As can be seen in section S3 of the SI, the $^{1}${A}$_{2g}$ state is not the only singlet state which has an energetically higher triplet counterpart, in violation of Hund{'}s rule. Another example is the $^{1}${A}$_{1u}$ state in cumulenic structure. It has disjoint SONOs, which are in the molecular plane rather than vertical to it. This analysis of orbitals underlines the importance of disjoint SONOs, which may shed light on designing new aromatic molecules. For completeness, we also visualized the SONOs of open-shell states of {C}$_{12}$ in section S4 of the SI. Clearly, the SONOs of the low-lying states of {C}$_{12}$ possess similar properties as those of {C}$_{8}$.

\subsection* {\label{PES} 2.4 Interconnections of polyynic and cumulenic states $\And$ mechanism of violation of Hund{'}s rule}

So far, we have discussed the polyynic and cumulenic states separately as being disjoint systems. In this section we plan to uncover how the low-lying states of one type relate to those of the other type. In the cumulenic geometry we find (see Figure 2) a singlet-triplet $^{1}${A}$_{2g}$-$^{3}${A}$_{2g}$ pair which violates Hund{'}s rule followed by a closed-shell $^{1}${A}$_{1g}$ state at higher energy. The global ground state is a closed-shell {1}$^{1}${A}$_{g}$ state and naively one anticipates that this polyynic state would become the closed-shell $^{1}${A}$_{1g}$ state when changing the geometry smoothly from polyynic to cumulenic. As we shall see below, this is not the case at all and the overall situation at hand is rather intriguing.

The first excited polyynic state is the triplet $^{3}${A}$_{g}$ state which indeed directly connects to the stable cumulenic triplet $^{3}${A}$_{2g}$ state. We computed the minimum energy path of the polyynic triplet state and found that it evolves to the lowest cumulenic triplet state which lies 0.75 eV lower in energy. The minimum energy path has been computed at the MRSDCI(12e, 11o)/cc-pVDZ level and is shown as red dots in Figure 4. The singlet counterpart of the polyynic $^{3}${A}$_{g}$ state is the {2}$^{1}${A}$_{g}$ state and this singlet-triplet pair exhibits a rather large singlet-triplet splitting of 1.1 eV explained in the preceding section. Again, one would naively expect this singlet state to directly evolve to the lower-lying cumulenic open-shell singlet $^{1}${A}$_{2g}$ which is a member of the $^{1}${A}$_{2g}$-$^{3}${A}$_{2g}$ singlet-triplet pair. And, again, this expectation is incorrect.

For notational ease, in the following paragraphs, we will use {S}$_{0}$ instead of {1}$^{1}${A}$_{g}$ state (in polyynic structure) and $^{1}${A}$_{2g}$ state (in cumulenic structure). Similarly, {T}$_{1}$ stands for the $^{3}${A}$_{g}$ state (in polyynic structure) and the $^{3}${A}$_{2g}$ state (in cumulenic structure), and {S}$_{1}$ stands for the {2}$^{1}${A}$_{g}$ state (in polyynic structure) and the $^{1}${A}$_{1g}$ state (in cumulenic structure).

As we know from the former section that the $^{1}${A}$_{2g}$ state of lowest energy in the cumulenic structure is a transition state, we have computed its minimum energy path at the same level as mentioned above. The path is depicted in Figure 4 and it is evident that the transition state in the cumulenic geometry evolves to become the global ground state, the {1}$^{1}${A}$_{g}$ polyynic state.  This minimum energy path of {S}$_{0}$ starts at the cumulenic geometry where {S}$_{0}$ is an open-shell state and ends at polyynic geometry where {S}$_{0}$ is a closed-shell state. The closed-shell character of the state, i.e., the weight of the closed-shell configuration in the MRCI wavefunction, is indicated in Figure 4 along the path. As seen in Figure 4, {S}$_{0}$ changes from a pure open-shell (0.000 closed-shell character) cumulenic state to a close to pure closed-shell polyynic state  (0.792 closed-shell character). While the character of the state is seen to change dramatically, the energy of {S}$_{0}$ changes only moderately by around 0.5 eV and, more importantly, the absolute changes of the geometry are rather small{:} the C-C bond lengths have changed by 0.06$\sim$0.07 $\AA$. The changes of the bond lengths indicated in Figure 4 along the path.

\begin{center}
\includegraphics[scale=0.5]{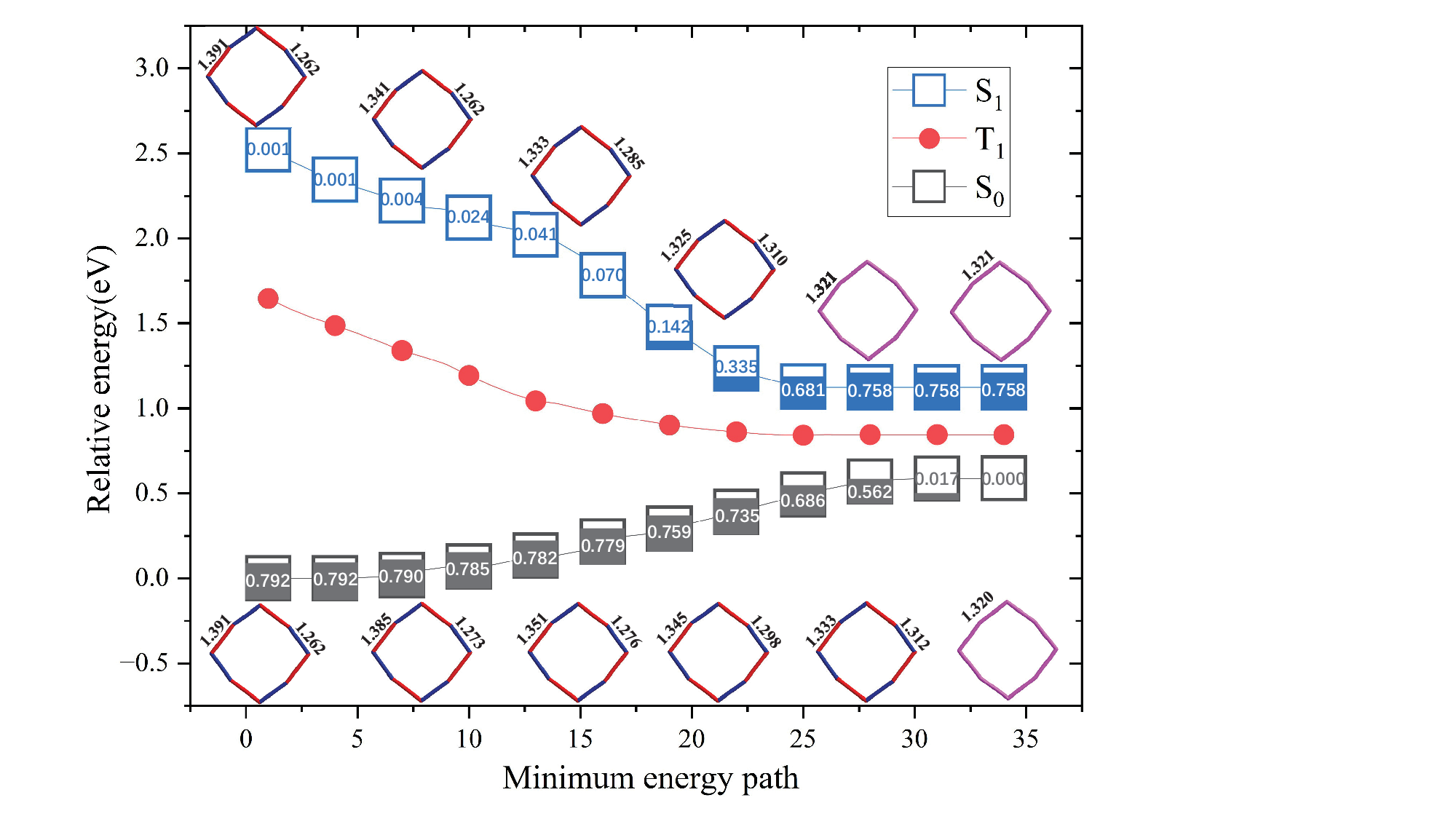}
\end{center}
\begin{flushleft}
{\bf Figure 4. The minimum energy paths of the three states ({S}$_{0}$ and {S}$_{1}$, triplet state {T}$_{1}$) computed at the MRCISD+Q(12e, 11o)/cc-pVDZ level of theory\cite{werner1988efficient,knowles1988efficient,knowles1992internally} employing Molpro program package.\cite{werner2012molpro}. The {S}$_{0}$, {T}$_{1}$, and {S}$_{1}$, states are marked in black, red and blue, respectively. The {T}$_{1}$ and {S}$_{1}$ states start the path from the equilibrium geometry of the ground state {S}$_{0}$ (polyynic), while the {S}$_{0}$ state starts from the equilibrium geometry of {T}$_{1}$ (cumulenic). Starting with the first optimization step of the minimum energy paths, we calculated the weight of the closed-shell configuration of {S}$_{0}$ and {S}$_{1}$ after every three optimization steps. The results are shown in the boxes depicted along the respective minimum path curves. Note the change of character in both minimum energy pathways of {S}$_{0}$ and {S}$_{1}$. The series of {C}$_{8}$ rings show the evolution of geometries along the minimum energy paths. The relative energies refer to the MRCI energies with respect to the equilibrium geometry of the global ground state, see Figure 1. The energy unit is eV.}
\end{flushleft}

For completeness, the minimum energy path of the {S}$_{1}$ state is also depicted in Figure 4. This path starts at the polyynic structure and drops by about 1.25 eV to the cumulenic structure. Again, the change in character of {S}$_{1}$ along the path is indicated in the figure. {S}$_{1}$  starts its path as a pure open-shell state (0.001 closed-shell character) to become a nearly closed-shell state (0.758 closed-shell character). The equilibrium geometry of the {S}$_{1}$ state in the cumulenic structure is very similar to the equilibrium geometry of {T}$_{1}$ state, and the bond length differs by only 0.001 $\AA$, at the MRCISD+Q/cc-pVDZ level.

Unfortunately, due to the high demand of the computational resources of the MRCI method, studying the minimum energy paths of {C}$_{12}$ is beyond our capability. However, all computations which we could perform show that the situation in the case of {C}$_{12}$ is very similar to that found for {C}$_{8}$.

\begin{center}
\includegraphics[scale=0.5]{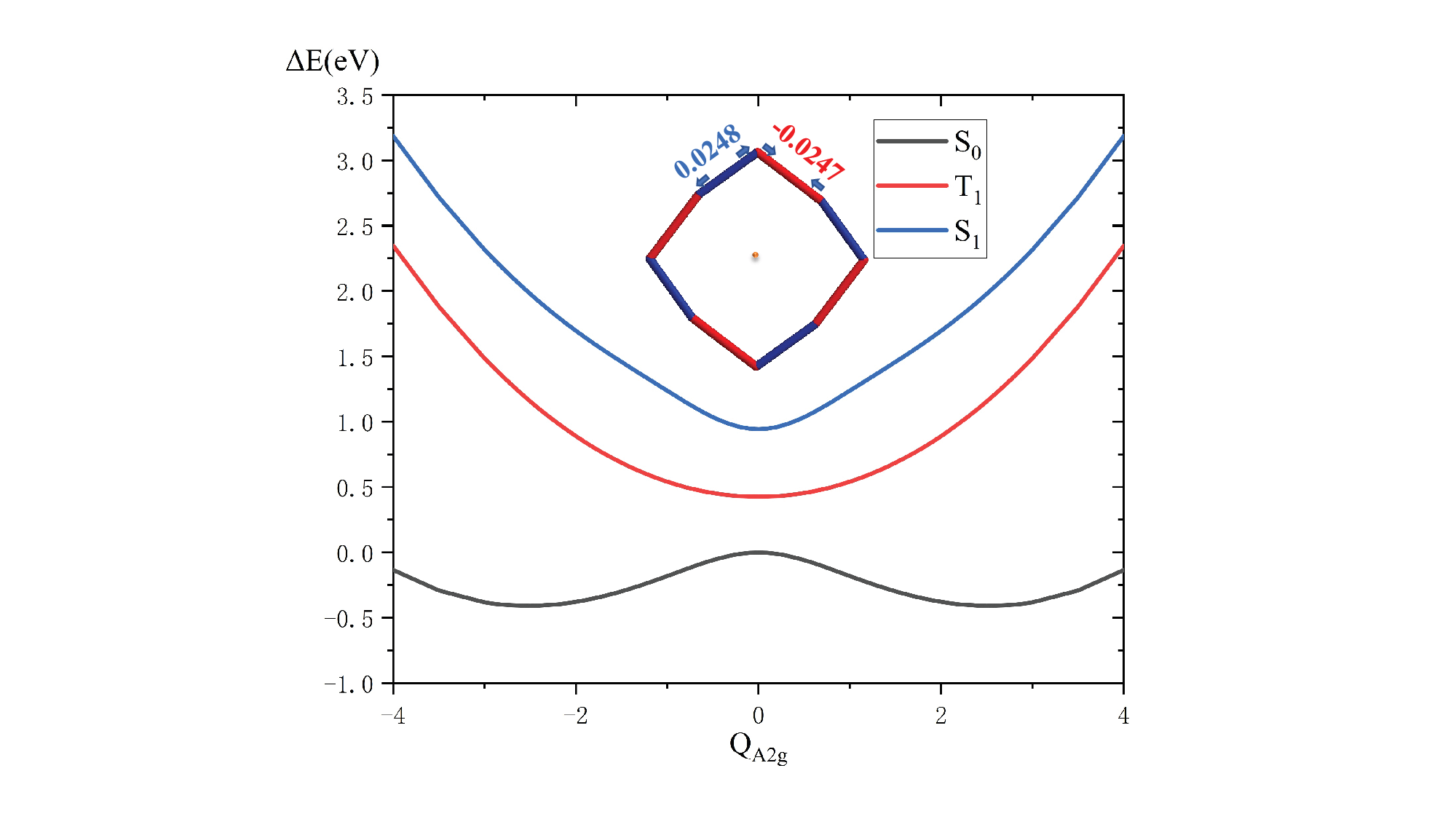}
\end{center}
\begin{flushleft}
{\bf Figure 5. Cuts through the potential energy surfaces of the three states {S}$_{0}$, {T}$_{1}$, and {S}$_{1}$ of the {C}$_{8}$ ring along the dimensionless vibrational normal mode {Q}$_{A2g}$ of the triplet state {T}$_{1}$. The cuts are depicted in black, red, and blue, respectively. At the equilibrium geometry of {T}$_{1}$, {Q}$_{A2g}${=0}. The {A}$_{2g}$ mode is envisaged in the inset, where the optimized structure of {T}$_{1}$ is shown and arrows indicate the changes of bond lengths due to the mode; the numbers relate to the changes of the bond lengths in $\AA$ at {Q}$_{A2g}${=1}. It is seen that the vibronic interaction between the {S}$_{1}$ and {S}$_{0}$ states leads to a symmetry breaking of the latter and the equal bond lengths at the cumulenic structure become distinguishable alternating. At the minimum of {S}$_{0}$ along the {Q}$_{A2g}$ mode, the energy of {S}$_{0}$ is lowered by 0.41 eV, simulating the optimized polyynic structure of {S}$_{0}$ as found in Figures 1 and 2.} 
\end{flushleft}

Let us now discuss the violation of Hund{'}s rule we have found for the $^{3}${A}$_{2g}$-$^{1}${A}$_{2g}$ pair at the cumulenic structure. We concentrate on the situation where these states are the lowest lying states. On the Hartree-Fock level of theory, the triplet state is lower in energy than its singlet partner by twice the exchange integral of the two singly occupied orbitals involved.\cite{helgaker2000molecular} Consequently, in general, a violation can be found if the singlet configuration is subject to a larger interaction with energetically higher lying excited singlet configurations than is the case for the triplet configuration, pushing the singlet state more down than the triplet one by at least twice as much as the value of the exchange integral. At the cumulenic structure we have seen in Figure 3 that the two singly occupied orbitals are distributed over alternating carbon atoms, such that their exchange integral can be expected to be very small. The triplet state has no close lying other triplet states (see Figure 2 and section S3 of the SI) with which it can interact. On the other hand, it is evident from Figure 4 that the two singlet states {S}$_{0}$  and {S}$_{1}$, or more precisely, the underlying leading electronic configurations, do interact substantially thereby exchanging their electronic character. As we can see in Figure 4, {S}$_{1}$ drops down steeply in energy along the path leading to the cumulenic structure while {S}$_{0}$ moves up to become a transition state and the two states come close to each other and can interact.

Although the mechanism of the found violation of Hund{'}s rule is now clarified, we can dive deeper into it as we notice that at the cumulenic structure the symmetry of {S}$_{0}$ is $^{1}${A}$_{2g}$ and that of {S}$_{1}$ is $^{1}${A}$_{1g}$. Along the paths shown in Figure 4, the two states have the same symmetry except at the cumulenic structure itself, where {T}$_{1}$ has its minimum. By applying the symmetry selection rules for the modes leading to linear vibronic coupling\cite{kouppel1984multimode} between the {S}$_{0}$ and {S}$_{1}$ states in the cumulenic structure, {D}$_{4h}$ point group,  one sees that the vibrational mode of {A}$_{2g}$ symmetry couples the two states breaking the symmetry.  Fortunately, there is only a single {A}$_{2g}$ vibrational mode in {C}$_{8}$. Consequently, we have computed the potential energy curves of {S}$_{0}$, {S}$_{1}$ and also of {T}$_{1}$ along the dimensionless vibrational normal mode Q$_{A2g}$ employing the EOM-SF-CCSD/cc-pVTZ method (see Methods section). The computed cuts through the potential energy surfaces along the {A}$_{2g}$ vibrational mode are shown in Figure 5.

As one can see from Figure 5, owing to the vibronic coupling mechanism, the symmetry breaks and the energy of the {S}$_{0}$ state decreases along the {A}$_{2g}$ mode. This nicely explains why {S}$_{0}$ is a transition state at the cumulenic structure.  At the minimum of the cut of the {S}$_{0}$ state, Q$_{A2g}$ is around 2.7. The corresponding bond length change with respect to the cumulenic structure is small (around 0.066 $\AA$). In contrast, the energy of the {S}$_{1}$ state increases along the {A}$_{2g}$ mode by the vibronic coupling. This pattern of symmetry breaking reflects a strong vibronic coupling between these two singlet states. Due to this coupling, the two singlet states interchange their closed-shell and open-shell characters as a function of the dimensionless vibrational mode.

\subsection* {\label{cumulenic} 2.5 How to arrive at the cumulenic structure{:} an excitation cycle}

The global ground state of {C}$_{8}$ is the {1}$^{1}${A}$_{g}$ state which possesses a polyynic structure. However, as shown in the preceding sections, the ring exhibits more interesting features at the cumulenic structure. In the following we briefly discuss how, starting from the polyynic {C}$_{8}$, we can arrive at the cumulenic structure. We propose here two possibilities utilizing the results shown in Figure 4. A, at first sight, straightforward way is to optically excite the {1}$^{1}${A}$_{g}$ ({S}$_{0}$) state to the {2}$^{1}${A}$_{g}$ ({S}$_{1}$) state as their energy difference is amenable to laser excitation. However, as the point group of the polyynic ring is {C}$_{4h}$, this transition is forbidden by optical selection rules via a dipole transition and only allowed by a quadrupole transition which is usually weak. Once the polyynic {S}$_{1}$ state is populated, nuclear dynamics will take it downhill to the cumulenic structure, see Figure 4. Here, unfortunately, we have the obstacle that due to numerical convergence problems, we could not find the absolute minimum of {S}$_{1}$ and this state at its cumulenic minimum might also turn out be a transition state as is $^{1}${A}$_{2g}$ ({S}$_{0}$).  The other possibility to access the cumulenic structure is via electron impact. It is well known\cite{falkowski2023benzene, ralphs2013water, jones2012pyrimidine, allan1989review} that triplet states can be well accessed from singlet states by electron impact. So, if one excites the polyynic {1}$^{1}${A}$_{g}$ ground state vertically to the $^{3}${A}$_{g}$ triplet state, we see in Figure 4 (where these states are named {S}$_{0}$ and {T}$_{1}$) that nuclear dynamics will take {T}$_{1}$ downhill along its minimum energy path to the cumulenic structure which is indeed its minimum on the global potential energy surface.

Having arrived at the inherently cumulenic {T}$_{1}$, this state can, in principle, be long-lived as the only way for this state to decay is by the mechanism of intersystem crossing\cite{gonzalez2020quantum} where the triplet {T}$_{1}$ crosses to the singlet {S}$_{0}$.  Intersystem crossing requires spin-orbit coupling to couple the {T}$_{1}$ and {S}$_{0}$ electronic states giving rise to phosphorescence or{/}and non-adiabatic coupling to enable the transfer of population. One may expect  the spin-orbit coupling to be weak for a carbon ring which would imply that the ring is long-lived in the cumulenic structure. If, on the other hand, intersystem crossing would take place, the {S}$_{0}$ state will follow its minimum energy path and the ring returns to its polyynic ground state completing the {S}$_{0}$ (polyynic) $\rightarrow$ {T}$_{1}$ (polyynic) $\rightarrow$ {T}$_{1}$ (cumulenic) $\rightarrow$ {S}$_{0}$ (cumulenic) $\rightarrow$ {S}$_{0}$ (polyynic) cycle. Following this cycle, will enable one to learn much about the intersystem crossing time and nuclear dynamics in the carbon ring. We hope that such an experiment and dedicated nuclear dynamics calculations will be possible one day.

\section{Conclusions}

In spite of the fact that Carbon rings have attracted much attention of both theory and experiment, little attention has been paid to their excited states or open-shell states. In this work, we systematically studied both the ground state and the excited states of {C}$_{8}$ and {C}$_{12}$ carbon rings. As expected, the global ground states have been found to be polyynic closed-shell structures. The lowest triplet states, on the other hand, are minima on the respective potential energy surface at cumulenic structures, indicating aromaticity. This highly accurate coupled cluster result is consistent with the prediction of Baird{'}s rule. By visualizing the singly occupied natural orbitals of the lowest singlet and triplet open-shell states at the computed cumulenic structures, we found that the two singly occupied natural orbitals are non-bonding and quasi-degenerate. Due to the disjoint nature of these SONOs, the lowest singlet and triplet open-shell states of {C}$_{8}$ and {C}$_{12}$ are disjoint diradicals. To the best of our knowledge, this is the first report on a stable diradical of carbon allotropes.

At the cumulenic structure which corresponds to the stable ground state of the triplet manifold, the partner singlet state has a lower energy and we encounter a violation of Hund{'}s multiplicity rule in both {C}$_{8}$ and {C}$_{12}$. This violation is analysed and can be attributed to the strong coupling between this singlet open-shell and the singlet closed-shell close by in energy at the cumulenic structure. This is the first report of a violation of Hund{'}s rule in carbon allotropes aside from graphenes.

To arrive at an understanding of the interconnection of the electronic states at the polyynic and those at the cumulenic structures, we have computed for {C}$_{8}$ (for {C}$_{12}$ the computational efforts are beyond reach for us) the minimum energy pathways of the two lowest in energy singlet states and of the lowest triplet state. Clearly, the polyynic triplet is seen to relax to its minimum which is the cumulenic triplet. The situation is more involved for the singlet states. While the first excited polyynic singlet is an open-shell state, it relaxes to the closed-shell singlet state at cumulenic geometry. The singlet open-shell state at the cumulenic geometry, which is the state lowest in energy at that geometry, is found to be a transition state which relaxes to the polyynic closed-shell state which is the global ground state of {C}$_{8}$. Applying the transparent symmetry selection rule of a linear vibronic coupling model for the two singlet states, one readily understands that the open-shell singlet is subject to symmetry breaking and becomes a transition state.

The differences in the geometry of the optimized polyynic and cumulenic structures are found to be rather small. The calculations including the vibronic coupling model explain why they nevertheless lead to prominent changes in the electronic structures. 

In their global ground state the rings are polyynic. On the other hand, the properties of the states in the cumulenic geometry are more diverse and hence seem more attractive. How to get from the polyynic ground state to the cumulenic triplet? Indeed, the found states and interconnecting pathways allow us to construct a scheme to arrive at the cumulenic structure of the ring. This scheme is discussed in section II.E. Briefly, starting from the polyynic ground state, one can vertically arrive at the polyynic lowest triplet by electron impact which is often rather efficient for singlet-triplet excitations. This triplet state relaxes to the cumulenic triplet where we expect it to be long-lived as it can only decay by intersystem crossing to the lowest singlet state and spin-orbit coupling is expected to be weak in carbon allotropes. Once it does decay to the singlet, this state relaxes to the polyynic ground state thus closing the cycle. This gives us the unique possibility to investigate the unusual intersystem crossing from a triple to a singlet due to the fact that Hund{'}s rule is violated.

We hope that the intriguing properties of carbon rings we have uncovered and analysed employing high-level methods, like large changes in electronic structure resulting upon small geometric changes, the unusual minimum energy pathways between polyynic and cumulenic structures, violation of Hund{'}s rule, Baird{'}s aromaticity and stable aromatic disjoint diradicals, enrich our knowledge of carbon allotropes and shed light on possible applications of carbon rings in the future.

\section{Methods}

Due to the contradiction encountered between the prediction of DFT and MP2 calculations and the experimental results for carbon rings mentioned in the introduction, it became evident that state-of-the-art high quality {\textit {ab initio}} methods should be employed for these systems. In this paper, we optimized the equilibrium geometries of the closed-shell states of {C}$_{8}$ and {C}$_{12}$ rings, employing the coupled cluster singles and doubles (CCSD)\cite{CCSD} method. To study the vertical excitation energies of singly excited states, the equation-of-motion coupled cluster method for electron excitations (EE-EOM-CCSD)\cite{stanton1993equation} has been used employing the Hartree-Fock wavefunctions of the lowest closed-shell state as reference functions. These aforementioned calculations are done employing CFOUR software package.\cite{cfour} To optimize the structures of low-lying excited states, we combined EOM-CCSD method with spin-flip method\cite{levchenko2004equation} based on the Hartree-Fock wavefunctions of the triplet state as reference functions, employing Q-chem program.\cite{Qchem} The basis set used in all coupled cluster calculations is Dunning{'}s correlation-consistent triple-zeta (cc-pVTZ)\cite{dunning1989gaussian, prascher2011a} basis set. The Cartesian coordinates of optimized geometries are shown in section S1 of the SI. 

To ensure that the optimized geometries of open-shell states provide minima on the energy surfaces, we have computed the respective harmonic vibrational frequencies. Thereby, we employed the CFOUR software package to generate symmetry displacements in all harmonic frequency calculations. It has great advantage that the energies computed using composite methods were determined by running multiple jobs at each displaced geometry and summing up the appropriate energies. In this case, the energy of each symmetry displacement was calculated employing the EOM-SF-CCSD method using the Q-chem program. Then, the final composite energies were communicated to CFOUR. A similar procedure was used in a recent article.\cite{franke2019ethyl} The results of vibrational frequencies are collected in section S2 of the SI. Thereby, the 1s core orbitals of carbon were kept frozen in all the aforementioned coupled cluster studies.

To compute the interconnection between states in the polyynic and cumulenic structures and to arrive at a better understanding of the mechanism leading to the violation of Hund{'}s rule we have found, we ran several minimum energy path calculations by employing the Multi-reference Configuration Interactions with Single and Double (MRCISD) method\cite{werner1988efficient,knowles1988efficient,knowles1992internally} and the generalized Davidson correction (+Q) in Molpro program package.\cite{werner2012molpro} Due to the limitation of computational resource, we concentrated on {C}$_{8}$ and used correlation-consistent double-zeta (cc-pVDZ)\cite{dunning1989gaussian, prascher2011a} basis set in MRCI calculation. Thereby, the inactive doubly occupied orbitals of carbon were kept frozen to save computational resource.

\section*{Acknowledgements}

The authors acknowledge the support by National Natural Science Foundation of China (Grants 12404325, 91961204, 11922405, 12274178, and 11874178). Assoc. Prof. Y.-F. Yang acknowledges Prof. Jochen Schirmer, Dr. Peter R. Franke, Prof. Devin Matthews, Prof. Dr. Frank Neese, Prof. Wenli Zou, Prof. Zikuan Wang, and Dr. Qi Song for valuable discussions. Prof. L. S. Cederbaum and Assoc. Prof. Y.-F. Yang mourn the sudden, untimely death of Prof. John F. Stanton, who gave so many valuable suggestions in coupled cluster calculations of this work.

\section*{Supporting information}

The Supporting Information provides additional computational details and data for the carbon rings studied in this work. It includes optimized Cartesian coordinates for the relevant polyynic and cumulenic structures of {C}$_{8}$ and {C}$_{12}$ , harmonic vibrational frequencies used to confirm the nature of the stationary points, and additional vertical excitation energies for low-lying singlet and triplet excited states. It also reports singly occupied natural orbitals and relative energies for the closed-shell and open-shell electronic states of polyynic and cumulenic {C}$_{8}$ and {C}$_{12}$.

\bibliography{Hund_C4n.bib}


\newpage

\rule{0.05in}{1.75in}%
\begin{minipage}[b][1.75in]{3.25in}
  \sffamily
  \includegraphics[scale=0.25]{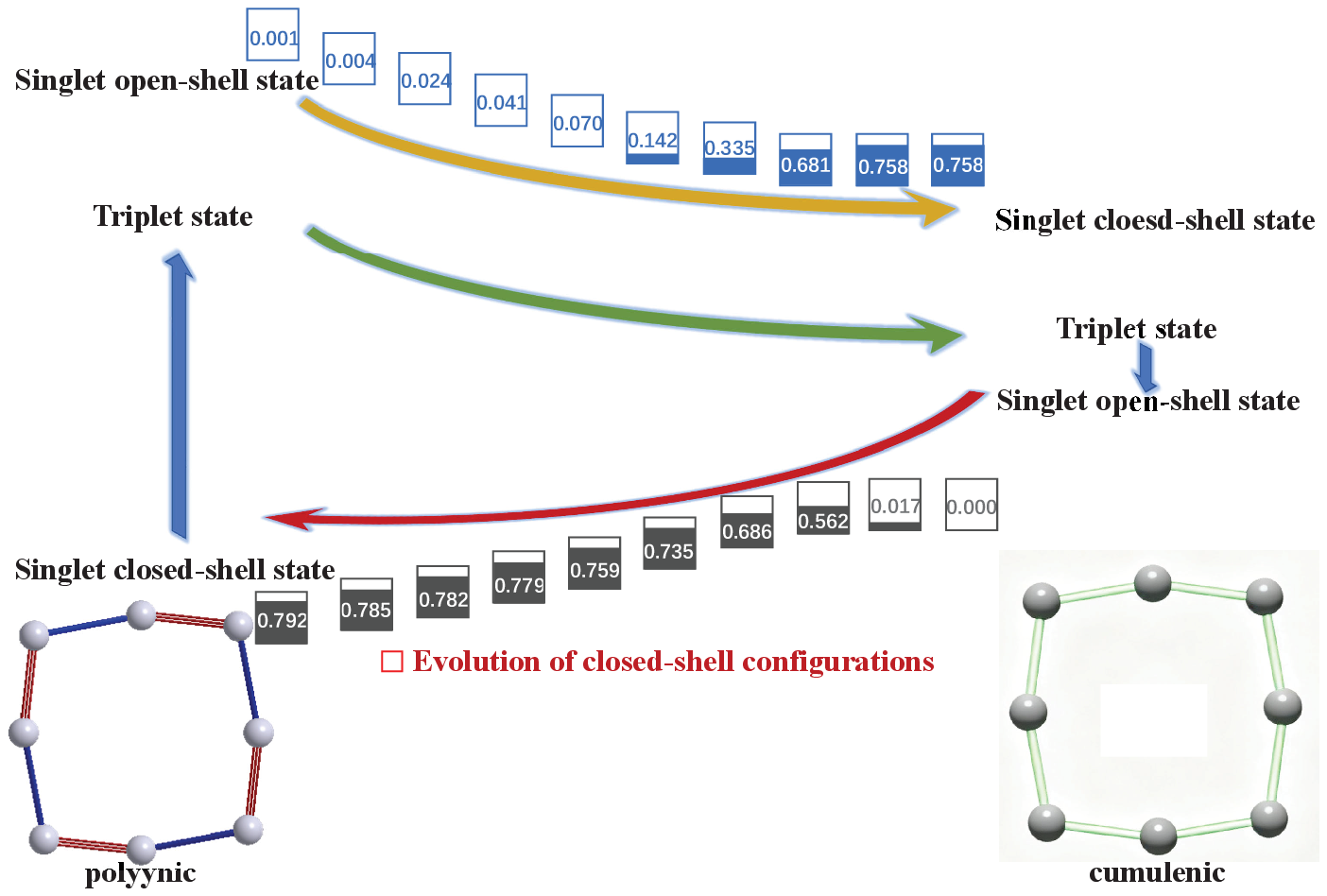}
  \frenchspacing

{\bf TOC Figure}

\end{minipage}%
\rule{0.05in}{1.75in}

\end{document}